\documentstyle[12pt]{article}

\newcommand{\be}{\begin{equation}}
\newcommand{\ee}{\end{equation}}

\newcommand{\bea}{\begin{eqnarray}}
\newcommand{\eea}{\end{eqnarray}}

\newcommand{\p}{\partial}

\newcommand{\nn}{\nonumber \\}
\newcommand{\f}{\frac}
\newcommand{\w}{\wedge}

\begin{document}
\thispagestyle{empty}

\begin{flushright}
{\bf arXiv: 1011.3117}
\end{flushright}
\begin{center} \noindent \Large \bf
Model building in AdS/CMT: DC conductivity and Hall angle
\end{center}

\bigskip\bigskip\bigskip
\vskip 0.5cm
\begin{center}
{ \normalsize \bf   Shesansu Sekhar Pal}

\vskip 0.5cm

\vskip 0.5 cm
Center for Quantum Spacetime, \\
Sogang University, 121-742 Seoul, South Korea\\
\vskip 0.5 cm
\sf shesansu${\frame{\shortstack{AT}}}$gmail.com
\end{center}
\centerline{\bf \small Abstract}

Using the bottom-up approach in a holographic setting, we attempt to study both the transport and thermodynamic properties of a generic system in $3+1$ dimensional bulk spacetime. We show the exact $1/T$ and $T^2$ dependence of the longitudinal conductivity and Hall angle,  as seen experimentally in most copper-oxide systems, which are believed to be close to quantum critical point. This particular temperature dependence of the conductivities are possible in two different cases: (1) Background solutions with scale invariant  and broken rotational symmetry, (2)  solutions with pseudo-scaling and unbroken rotational symmetry but only at low density limit.
Generically, the study of the transport properties in a scale invariant background solution, using the probe brane approach,  at high density and at low temperature limit suggests us to consider only metrics with two exponents. More precisely, the spatial part of the metric components should not be  same i.e., $g_{xx}\neq g_{yy}$. In doing so, we have generalized the above mentioned behavior of conductivity with a very special behavior of specific heat
which at low temperature goes as: $C_V\sim T^3$. However, 
if we break the scaling symmetry of the background solution   by  including  a nontrivial dilaton, axion or both and keep the rotational symmetry  then 
also we can generate such a behavior of conductivity but only in the low density regime.
As far as we are aware, this particular temperature dependence of both the conductivity and Hall angle is being shown for the first time using holography.
\newpage
\tableofcontents
\section{Introduction}

There are interesting model building calculations that are being put forward using gauge/gravity duality, which suggests to have captured the experimental  results close to quantum criticality and the associated  quantum phase transitions. In particular, for the copper-oxide systems  at low temperature, the resistivity, which is the inverse of the conductivity, goes as $\sigma\sim T^{-1}$ \cite{cwo},~\cite{pwa},~\cite{tm},~\cite{rc}. 
This interesting behavior has been reported  in a controllable  yet unrealistic setting for a very special kind of gravitational system  that  displays the Lifshitz like property and is possible only when the Lifshitz exponent takes a special value namely, $z=2$, \footnote{There is another paper \cite{filmv}, which does not require gravitational solution with Lifshitz scaling (rather with $z=1)$ in order to generate such a behavior of conductivity. More interestingly, it is shown that such a behavior  follows at one loop.} \cite{hpst}. However, it is also suggested in  \cite{cwo},~\cite{pwa},~\cite{tm},~\cite{rc} that for the copper-oxide systems, the Hall angle,  $cot~\theta_H=\sigma^{xx}/\sigma^{xy}$, should have a quadratic dependence of temperature, $cot~\theta_H\sim T^2$. But, unfortunately, use of the gravitational solutions showing the Lifshitz like scaling does not reproduce this behavior of Hall angle, rather it gives at low temperature a linear dependence of temperature and is not in complete agreement with the experimental results.

The experimental results for the transport properties of the copper-oxide systems near optimum doping at low temperature can be summarized as follows \cite{cwo},~\cite{tm},~\cite{rc}
\be\label{expt_result}
\sigma^{xx}\sim 1/T,~~~cot~\theta_H=\sigma^{xx}/\sigma^{xy}\sim T^2~~~\Longrightarrow ~~~ \sigma^{xy}\sim T^{-3}.
\ee

The basic reason of not getting the desired experimental behavior is due to the presence of a rotational symmetry  in the $x,~y$ plane of the metric while having the scaling symmetry of the background solution, where  $x$ and $y$ are the only two spatial directions available in field theory. Even though this symmetry is broken explicitly  in the presence of constant electric and magnetic field. 

In this paper we shall show that  eq(\ref{expt_result}) can  only be reproduced  in two different cases  (1) background solutions respecting the scaling symmetry  with broken rotational symmetry in the $x,~y$ plane  (2) pseudo-scaling background solutions with unbroken rotational symmetry in the low density limit. Here the  pseudo-scaling  solutions means, the background geometry respects the scaling symmetry but not the scalar fields like dilaton and axion. Furthermore, the background solutions having  the scaling symmetry, time translation, spatial translation, and the rotational symmetry are completely ruled out by eq(\ref{expt_result}), e.g.,  pure  AdS and pure Lifshitz solutions. It is worth to emphasize that case (1) is the only choice that is permissible at high density, but at low density we can have either of the choices.   We are discussing both the limits of densities because it is not {\it a priori} clear the scale of optimum doping in eq(\ref{expt_result}).

The basic philosophy of \cite{hpst} is to introduce charge carriers via Dp branes, in the probe brane approximation. The charge carriers  are in thermal contact with a heat bath, which is taken as the Lifshitz black hole.  Translating it into the language of \cite{kk}, it's the bi-fundamental degrees of freedom that are charged,  interacting among themselves and with the adjoint degrees of freedom  giving us the desired feature of conductivity. In contrast to \cite{filmv}, where the authors have considered only the charged adjoint degrees of freedom to replicate the above mentioned experimental result at one loop\footnote{There arises a natural question: Is this behavior of charged bi-fundamental degrees of freedom in a heat bath=1-loop adjoint degrees of freedom in a different heat bath,  generically ? Which we are not going to address.}. In this paper, we have adopted the former approach (in the massless limit) and replace the heat bath of Lifshitz kind by another, more general, heat bath. The reason of such a replacement is that (1) Lifshitz type heat bath is a special type to this more general heat bath (2) It's the    eq(\ref{expt_result}), which was not possible to reproduce fully with the Lifshitz type heat bath. Recall,  the heat baths are essentially the source of  studying physics around  the quantum critical point at low temperature \cite{qpt}.   The consequences of replacing such a heat bath is  addressed, thermodynamically. 

In the holographic setting \cite{jm}, the  authors   of  \cite{kob},~ \cite{ob} and \cite{ks}  have  proposed a beautiful algorithm  to calculate the conductivities. Here we have modified it slightly and obtain an equivalent way to calculate the conductivities. The result of the calculation matches precisely as is done in \cite{ob} 
when the charge carriers move in a constant electric and magnetic field.  Use of this equivalent prescription leads to the following dependence of  conductivities on the metric components evaluated at some holographic energy scale,  $r_{\star}$. At high densities compared to temperature 
\be\label{conductivities_no_cs}
\sigma^{xx}\sim  \f{c_{\phi}e^{-2\Phi(r_{\star})}}{g_{xx}(r_{\star})},~~~\sigma^{xy}\simeq \f{B c_{\phi}e^{-4\Phi(r_{\star})}}{g_{xx}(r_{\star})g_{yy}(r_{\star})},
\ee
where $c_{\phi}$ is the charge density,  $B$, the magnetic field, and $\Phi$, the dilaton. In eq(\ref{conductivities_no_cs}), the spatial parts of the metric components along $x$ and $y$ directions are denoted as $g_{xx}$ and $g_{yy}$, respectively. Note, this result follows  when  the probe brane action admits only the DBI type of action. If we do include the Chern-Simon part of the action to the probe brane as well then the result of the  conductivities  gets slightly modified in the   high density limit compared to temperature 
\be\label{conductivities_with_cs}
\sigma^{xx}\sim  \f{c_{\phi}e^{-2\Phi(r_{\star})}}{g_{xx}(r_{\star})},~~~\sigma^{xy}\simeq \f{B c_{\phi}e^{-4\Phi(r_{\star})}}{g_{xx}(r_{\star})g_{yy}(r_{\star})}-\mu C_0(r_{\star}),
\ee
where $\mu$ is the coupling of the Chern-Simon action and $C_0$ is the axion field. Note that the scaling symmetry  is broken for a non constant dilaton and axion field. 

Now if we restrict ourselves to background solutions which possesses the scaling symmetry and    exhibits the rotational symmetry at the level of metric not the full system, then the off diagonal part of the conductivity in the   high density limit goes as,  $\sigma^{xy}~\sim~\bigg(\sigma^{xx}\bigg)^2$, which is not  in accordance with the experimental result, see eq(\ref{expt_result}). This means to reproduce  eq(\ref{expt_result}), in the high density limit we are  forced  to consider metric components for which $g_{xx}\neq g_{yy}$. This is one of the basic criteria that must be imposed in choosing the background metric, i.e.,  the heat bath,  in order to study the physics associated to the transport properties around the quantum critical point.

In getting the results of the conductivity and the Hall angle as in eq(\ref{expt_result}), we have assumed the background metric  respect the following scaling symmetry
\be\label{two_exponent_scaling}
t\rightarrow \lambda^z ~t,~~~x\rightarrow \lambda^w ~x,~~~y\rightarrow \lambda ~y,~~~r\rightarrow \f{r}{\lambda},
\ee
and also we have assumed that there is not any non trivial  scalar field like dilaton or axion, in the entire set up. As the presence of  such a non trivial background field would give rise to some-kind of pseudo-scaling theory. Of course, the  charge density of the bi-fundamental degrees of freedom, i.e.,
two form field strength, $F_2$, that appear in the DBI action, breaks the scaling symmetry.
More exactly, the gravitational solution without any non vanishing scalar field  that give us the desired result of the longitudinal conductivity and Hall angle should have the exponents, $z=1$ and $w=1/2$. For this choice of exponents, the zero temperature limit  of the black hole solution, i.e., the solution without the thermal factor, has a boost symmetry along the $t,~y$ plane with a form
\be
ds^2=L^2[-r^2dt^2+rdx^2+r^2dy^2+\f{dr^2}{r^2}],
\ee
where $L$ is the size of the $3+1$ dimensional bulk system, which we shall set to unity in our calculations latter. 
The background geometry with two exponents $z$ and $w$ 
was proposed in \cite{ssp} using a combination of Einstein-Hilbert action  and several form field strengths. Since, the analytic non-extremal version of that solution is very difficult to obtain. So, here we have adopted a different path to generate such a solution by using only gravitons.

Let us do a little bit of  dimensional analysis of various  physical quantities. If the $d+1$ dimensional field theory spacetime coordinates (i.e., the bulk is $d+2$ dimensional spacetime) behaves under scaling as
\be
t~\rightarrow~ \lambda^z ~t,~~~x~\rightarrow~\lambda^w~x, ~~~y_i~\rightarrow~ \lambda~y_i,~~~(i=1,\cdots,d-1) 
\ee

then  the physical quantities possesses the following length dimension 
\bea
&&[t]=z,~[x]=
w,~[y_i]=1,~[J^t]=1-d-w,~[J^x]=1-z-d,~[J^i]=2-z-d-w\nn
&&[A_t]=-z,~[A_x]=-w,~
~[A_i]=-1, ~[E_x]=-w-z,~
 [E_i]=-1-z,~[B_x]=-2,\nn
&& [B_i]=-1-w,~[T]=-z=[\omega],~[F]=-z,~[\sigma^{xx}]=1+w-d,
~[\sigma^{xy}]=2-d,\nn&&[\sigma^{yy}]=3-d-w,
\eea 
where $J^t,~J_i,~A_t,~A_x,~A_i,~E,~B,~T,~\omega,~F,~\sigma$ are charge density, current density, time component of the gauge 
potential, x-component of the gauge 
potential, $y_i$-component of the gauge potential, electric field, magnetic field, temperature, frequency,  free energy and conductivity, respectively. The two form field strength has the following form i.e., $F_2=-E_x dt\w dx-E_{y_i} dt\w dy_i+B_{y_j} dx\w dy_i+B_x dy_i\w dy_j+\cdots$.

In the small magnetic field and at low density limit, $c^2_{\phi}\ll {\cal N}^2 e^{2\Phi}g_{xx}g_{yy}$ but with $c_{\phi} \gg B\mu C_0$,  the conductivities are
\be
\sigma^{xx}\sim {\cal N}  e^{-\Phi(r_{\star})} \sqrt{\f{g_{yy}(r_{\star})}{ g_{xx}(r_{\star})}},~~~\sigma^{xy}\sim \f{B  c_{\phi}e^{-4\Phi(r_{\star})}}{ g_{xx}(r_{\star}) g_{yy}(r_{\star})}-\mu C_0(r_{\star}),
\ee
where ${\cal N}$ is the effective  tension of the brane.
Upon comparing with eq(\ref{expt_result}), 
we can generate the desired experimental behavior of transport quantities for background solutions showing the  
pseudo-scaling symmetry and unbroken rotational symmetry in the $x,~y$ plane. So, only in the low density limit we need not have to consider two exponents solution as in eq(\ref{two_exponent_scaling}). However, if we do then we can as well generate eq(\ref{expt_result}). 

In summary, the different possibilities with time translation and spatial translation symmetries are:
\be\label{summary_eq_1}
    \begin{tabular}{ | l | l | l |l|}
    \hline
 Symmetries & Density & Limit &Eq(1)\\
\hline\hline
Scaling and rotation&Any density & &Not possible \\ \hline\hline
Pseudo-scaling & Medium to & $c_{\phi} \gg \mu B C_0$&Possible  \\
and rotation & low density& $c_{\phi}\ll {\cal N} e^{\Phi}g^2_{xx}$  &  \\ \hline
Pseudo-scaling & High density & & Not possible  \\
and rotation & &  &  \\ \hline
Pseudo-scaling & Medium to  & $\f{Bc_{\phi}e^{-4\Phi}}{g^2_{xx}} \ll \mu  C_0$&  possible  \\
and rotation & low density & $c_{\phi} \gg \mu B C_0$ &  \\ \hline\hline
Scaling with broken rotation & Low density & &Possible\\\hline
Scaling with broken rotation & High density & &Possible\\
 \hline\hline
    \end{tabular}
\ee

The holographic study of the transport properties using the approach of \cite{kob},~ \cite{ob} and \cite{ks}  gives us the non-linear behavior  at the critical point and help us to understand the universal features, if any, in different limits of the parameter space, especially  the quantity $dI/dV=1/{\cal R}=\sigma$, where ${\cal R}$ is the resistance to the flow of current $I$ with an applied voltage $V$. In this paper, we have generated successfully eq(\ref{expt_result}),  and focused more on the model building than trying to find  the universal features.

In the calculation of the conductivity, it is not {\it a priori}  clear at what scale one should   evaluate, i.e., how to choose the scale,  $r_{\star}$,  so as to capture the non-linear effect.  Especially, for the system that is described by the Maxwell action. Of course, the gauge/gravity duality suggests us to do the calculations at the UV boundary. But, the  
result of this calculation produces only the linearized effect. However, for the system whose action is described by  the 
DBI type there exists a very natural way to find the scale  $r_{\star}$. This basically follows from the argument of \cite{kob},\cite{ob} and \cite{ks}, which says that either  the integrand of the action or the solution, which is in the form of $\sqrt{\f{A}{B}}$ needed to be real.  At a special value of the radial coordinate, $r=r_{\star}$, both $A$ and $B$ vanishes and there the action and the solution takes an indeterminate, $\f{0}{0}$ form. Above or below this special scale, $r_{\star}$, both $A$ and $B$ becomes positive or negative together. In this paper we give a physical argument to determine the scale $r_{\star}$ and show that it agrees precisely with the calculations done using the arguments of \cite{kob},\cite{ob} and \cite{ks}. We use the fact that the  Legendre transformed action  is same as the energy density, $H_L$, evaluated on  the static solution, which comes as the square root of one term, importantly there is not any term in the denominator. 
On this energy, we use the argument of \cite{kob},\cite{ob} or \cite{ks} to find the scale, $r_{\star}$.  So, the scale, $r_{\star}$, is the point on the holographic direction, $r$, for which the Legendre transformed action or the energy density vanishes and stay real
\be\label{condition_vanishing_energy_static}
\bigg(H_L\bigg)_{r_{\star}}=0.
\ee

For the systems that are described by DBI kind of actions, there exists another argument that precisely give the same result for $r_{\star}$ as suggested in the previous paragraph, even though the precise physical reason  is not that clear. The argument is to find the on shell value of the norm of the field strength  for which it takes a constant value, more precisely 
\be\label{condition_r_star1}
\bigg(F_{MN}F^{MN}\bigg)_{r_{\star}}=-2. 
\ee

There exists yet another way to determine the scale,  $r_{\star}$,  that is to find a  scale where the determinant of $det(g+F)_{ab}$ vanishes \cite{ob}. Here, the indices $a$ and $b$ run, only, over the field theory directions. The equation for the condition is
\be\label{condition_r_star2}
\bigg(det(g+F)_{ab}\bigg)_{r_{\star}}=0.
\ee
This can  very easily be  seen following  the argument of vanishing  of $H_L$ at the scale $r_{\star}$. Generically the  Legendre transformed action can be written as
\be\label{energy_legendre_transformed_generic}
H_L=\int \sqrt{{\cal A}(r)[{\cal A_2}(r){\cal A_3}(r)-({\cal A_4}(r))^2]},
\ee
where ${\cal A}(r)=\sqrt{\f{g_{rr}}{g_{tt}g_{xx}g_{yy}}}$ and the expression to ${\cal A_2}(r),~{\cal A_3}(r),~{\cal A_4}(r)$ are given in eq(\ref{A_2-A_3-A_4}). 
Generically, the term $({\cal A_4}(r))^2$ is  non-zero when there exists  more than one spatial current, more importantly this term is always positive. Whereas the term ${\cal A_2}(r)$ and ${\cal A_3}(r)$ can change sign close to the horizon. Hence, we can use the arguments of \cite{kob},  \cite{ob} and \cite{ks} so as to have a real Legendre transformed action or the energy. Moreover, one of the term  is nothing but  $(-det(g+F)_{ab})$. Hence,  the condition, eq(\ref{condition_r_star2}), follows from $H_L$.

The prescription of holography \cite{jm} or that of \cite{kob}   has been used to calculate the conductivity of several systems both in the top-down and bottom-up approaches. They include \cite{hkss},--,\cite{bbdl} as a partial list. 

This paper is organized as follows. In section 2, we shall review the calculation of the conductivity following \cite{kob} and compare it with that  using eq(\ref{energy_legendre_transformed_generic}) for systems that are described by DBI type of actions but in the absence of the  charge density. In section 3, we study the  system  in the presence of charge density  and with    Chern-Simon type of actions. Studies of section 2 and 3 are done in generic background solutions. Based on the calculations given in section 3, we give a toy example which is modeled in such a way that it gives us the desired behavior of conductivity and Hall angle, in section 4. In section 5, we study the thermodynamics of the charge carriers in the presence of a constant magnetic field. Finally, we conclude in section 6. Several details of the calculations are relegated to Appendices.

\section{From non-linear DBI action}

In this section, we shall evaluate the on-shell value of the  current using the definition,  $J^{\mu}=\f{\delta S}{\delta A_{\mu}}$. In  arbitrary  spacetime dimensions, it is very difficult to solve the equations of motion that results from the DBI action, even in the massless and zero condensate limit, i.e., for trivial embedding functions. Here, for simplicity, we shall restrict ourselves to $3+1$ dimensional bulk  spacetime.

The DBI action is
\be
S_{DBI}=-T\int e^{-\phi}\sqrt{-det([g]_{ab}+F_{ab})}\equiv
-T\int e^{-\phi}\sqrt{-det(M_{ab})},
\ee
where $[~]$ is used to denote the pull back of the bulk metric onto the world volume of the brane and $T$ is the tension of the brane. For simplicity, we have dropped the Chern-Simon part of the action.

Looking at the existence of an exact solution to Maxwell system in $3+1$ dimensions, as shown in Appendix A, suggests there could be an exact solution to the non-linearly generalized Maxwell system that is the DBI action. 

Let us assume the following structure to the metric and U(1) gauge field strength
\bea
ds^2_4&=&-h(r) d\tau^2+2d\tau dr+e^{2s(r)}(dx^2+dy^2),\nn 
F_2&=&F_{r\tau}dr\w d\tau+F_{x\tau}dx\w d\tau+F_{y\tau}dy\w d\tau+F_{xy}dx\w dy+F_{xr}dx\w dr+F_{yr}dy\w dr \nn
\eea

The equation of motion to gauge field and the  current associated to it are
\be
\p_{K}[T~e^{-\phi}\sqrt{-det(M_{AB})}~~~ \theta^{KL}]=0,~~~
 J^{\mu}=-T~e^{-\phi}\sqrt{-det(M_{AB})}~~~ \theta^{r\mu},
\ee
where the indices $N,~K,~L$ etc run over the entire bulk spacetime whereas $\mu,~\nu,~\rho$ etc run only over  the field theory directions, $\tau,~x,~y$. The function $\theta^{KL}=\f{M^{KL}-M^{LK}}{2}$ and the inverse of matrix, $M_{KL}$, is defined as,  $M^{KL}M_{LP}=\delta^K_P$.
The explicit form of the spatial components of the current are
\bea
\sqrt{-det(M_{AB})}~~~J^x&=&-T~e^{-\phi}\bigg[F_{y\tau}(F_{r\tau}F_{xy}+F_{x\tau}F_{yr}-F_{xr}F_{y\tau})+e^{2s}(F_{x\tau}+hF_{xr})\bigg],\nn
\sqrt{-det(M_{AB})}~~~J^y&=&-T~e^{-\phi}\bigg[F_{x\tau}(-F_{r\tau}F_{xy}+F_{y\tau}F_{xr}-F_{yr}F_{x\tau})+e^{2s}(F_{y\tau}+hF_{yr})\bigg]
\eea

Let us assume that the non-vanishing components of the field strengths are $F_{x\tau},~F_{y\tau}$  and $F_{xy}$. In which case
there occurs a lot of simplification to both the currents and  equations of motion 
\bea
&&J^x=-T~e^{-\phi}\f{F_{x\tau}}{\sqrt{1+e^{-4s}F^2_{xy}}},~~~J^y=-T~e^{-\phi}\f{F_{y\tau}}{\sqrt{1+e^{-4s}F^2_{xy}}},\nn &&
\p_y\bigg[T\f{e^{-\phi}~F_{xy}}{\sqrt{e^{4s}+F^2_{xy}}}\bigg]+\p_r\bigg[T\f{e^{-\phi+2s}~F_{x\tau}}{\sqrt{e^{4s}+F^2_{xy}}}\bigg]=0,~~~
\p_x\bigg[-T\f{e^{-\phi}~F_{xy}}{\sqrt{e^{4s}+F^2_{xy}}}\bigg]+
\p_r\bigg[T\f{e^{-\phi+2s}~F_{y\tau}}{\sqrt{e^{4s}+F^2_{xy}}}\bigg]=0,\nn &&\p_x\bigg[Te^{-\phi+2s}\f{F_{x\tau}}{\sqrt{e^{4s}+F^2_{xy}}}\bigg]+\p_y\bigg[Te^{-\phi+2s}\f{F_{y\tau}}{\sqrt{e^{4s}+F^2_{xy}}}\bigg]=0.
\eea

The solution for $\phi={\rm constant}=\phi_0$ becomes 
\be
F_{y\tau}=E_y(\tau,~r),~~~
F_{xy}={\rm constant}\equiv B ,~~~F_{x\tau}=E_x(\tau,~r),
\ee
for some functions $E_x(\tau,~r)$ and $E_y(\tau,~r)$, whose functional form is 
\be
E_x(\tau,~r)=f_2(\tau) e^{-2s}\sqrt{e^{4s}+B^2},~~~E_y(\tau,~r)=f_3(\tau) e^{-2s}\sqrt{e^{4s}+B^2}
\ee
determined in terms of two unknown functions $f_2(\tau)$ and
$f_3(\tau)$. The Bianchi identity sets the condition on $F_{x\tau}$ and $F_{y\tau}$ that these components should not depend on $r$ and  can happen only when $B=0$. This solution indeed is an exact solution  to the complete equation of motion and the only non-vanishing components of field strength are  $F_{x\tau}$  and $F_{y\tau}$ \cite{ks}.\\

Using this explicit structure of the solution in the expression of currents, we ended up with
\be\label{currents_dbi_4d}
J^x=-T~e^{-\phi}~F_{x\tau},~~~J^y=-T~e^{-\phi}~F_{y\tau},
\ee  
from which there follows  the DC conductivities at the scale,  $r=r_c$, upon using the Ohm's law  
\be
\sigma^{xx}(r_c)=\sigma^{yy}(r_c)=-T~e^{-\phi_0}\equiv\sigma.
\ee 

This indeed reproduces the result of \cite{ks}. There exists another exact solution but unfortunately with zero electric field and the non-vanishing components of the field strengths  are
\be
F_{xy}=B={\rm constant},~~~F_{r\tau}=\f{f_1}{\sqrt{f^2_1+e^{4s}+B^2}},
\ee
where $f_1$ is a constant.

\subsection{Comparing with the approach of \cite{kob}}

In this  subsection, we shall try to derive the expression of the conductivity from the  DBI action in the absence of density. Let us work in a $d+2$ dimensional spacetime with  dynamical exponent, $z$. The exact form of the metric that we shall be considering is
\be\label{lifshitz_geometry}
ds^2_{d+2}=-r^{2z}f(r) dt^2+r^2\sum^{d}_{i=1}dx^2_i+\f{dr^2}{r^2f(r)},
\ee 
 where we shall take $f=1-(r_0/r)^{d+z}$. This from of the metric gives us the Hawking temperature, $T_H=\bigg(\f{d+z}{4\pi}\bigg)r^z_0$, where $r_0$ is the horizon. In order to carry out the analysis for conductivity, we need to turn on a U(1) gauge potential which will give us the desired electric field in the field theory and for convenience we shall consider it as  constant. Along with this, we shall turn on another  component of the field strength, whose one leg is along the radial direction and the other along the spatial direction. For specificity, we shall turn on  $F_{xr}$. So, the complete form of the  U(1) gauge field is $F_2=-E dt\w dx-H'(r)dr\w dx.$

Let us consider a probe brane which is extended along time (t), radial direction (r) and $d_s-1$ number of directions of the $d$ number of spatial directions. Hence the probe brane is a $d_s$ brane. For $d_s=d+1$, the probe brane is a space filling brane. Also, for simplicity, we shall consider the  massless limit scenario and the action  becomes
\be\label{action_no_density}
S=-N\int dt dr dx d^{d-1} y \sqrt{\prod^{d-1}_{1}g_{y_ay_a}}\sqrt{g_{tt} g_{rr}g_{xx}+H'^2 g_{tt}-E^2 g_{rr}},
\ee
where we have considered the metric to be far more general than that written in eq(\ref{lifshitz_geometry}) but assumed to be  diagonal\footnote{Note, this form of the metric can very easily be re-written like that written in eq(\ref{EF-metric}), i.e., by doing a coordinate transformation.}. The explicit form of the metric that we have considered has the following structure
\be\label{generic_metric}
ds^2_{d+2}=-g_{tt}(r)dr^2+g_{rr}(r)dr^2+g_{xx}(r)dx^2+
\sum^{d-1}_{1}g_{ab}(r)dy^ady^b,
\ee
where $\sum^{d-1}_{1}g_{ab}(r)dy^ady^b$ is assumed to be diagonal too, i.e., $\sum^{d-1}_{1}g_{ab}(r)dy^ady^b=g_{11}(dy^1)^2+\cdots+g_{d-1,d-1}(dy^{d-1})^2$. 
The normalization $N$ includes the tension  and the number of  the probe branes. Since the action eq(\ref{action_no_density}) does not depend on the function $H(r)$ means  the `momentum' associated to it must be a constant i.e., $\f{\delta S}{\delta H'}\equiv c$. From which there follows  the solution
\be
H'=\pm c \sqrt{\f{g_{rr}g_{tt}g_{xx}-E^2 g_{rr}}{N^2(\prod_a g_{y_ay_a})g^2_{tt}-c^2 g_{tt}}}.
\ee

It is very easy to convince oneself that the constant, $c$, is nothing but the current density, $J^x$. Now, using the arguments of \cite{kob}, we obtain the  necessary equations to fix $c$, which is $J^x$, as
\be\label{conditions_no_density}
E^2=g_{tt}(r_{\star})g_{xx}(r_{\star}),~~~
J^2_x=N^2\bigg(\prod_a g_{y_ay_a}(r_{\star})\bigg) g_{tt}(r_{\star}),
\ee 
where $r_{\star}$ is the value of $r$, where both the numerator and denominator of $H'$ changes sign. It is interesting to note that at, $r_{\star}$, the gradient of the solution takes  $H'=\f{0}{0}$ form, which is an indeterminate structure. So,  the better way to find $r_{\star}$ is to go  over to an equivalent form  and demand that the energy density (or the Legendre transformed action) that follows  is real as well as  have a `minimum' at some energy scale, which we denote it as $r_{\star}$, too.

The action eq(\ref{action_no_density}) can equivalently be expressed
by doing the Legendre transformation as
\be
S_L=S-\int \f{\delta S}{\delta H'} H'=-\int \sqrt{\bigg[g_{rr}g_{tt}g_{xx}-E^2 g_{rr}\bigg]\bigg[N^2(\prod_a g_{y_ay_a})-\f{c^2}{g_{tt}} \bigg]}.
\ee
For static configuration, the  energy of the system is, $H_L=-S_L$, where  

\be\label{energy_electric_field}
H_L=\int \sqrt{\bigg[g_{rr}(g_{tt}g_{xx}-E^2) \bigg]\bigg[N^2(\prod_a g_{y_ay_a})-\f{c^2}{g_{tt}} \bigg]}.
\ee
  
For an illustration, let us take  an example of the asymptotically  AdS black hole, the first term in the square bracket  under the square root changes sign some-where close to horizon and the same is true for the second term in the square bracket and their product is positive. Since both the terms in the square bracket changes sign some-where close to the horizon, we assume that this happens at the same  value of radial coordinate, $r=r_{\star}$, so as to have a real energy or real Legendre transformed action. Asymptotically, the first term in the square bracket diverges so also the second term (for $d \geq 2$). Now the only place it can vanish (i.e., minimum) is close to the horizon. For a discussion on the condition of minimization to energy, see  Appendix B.

Demanding these two restrictions  again gives us the same two equations as  written in eq(\ref{conditions_no_density}). From which  there follows the expression of current 
\be
J^x=\pm N\f{\sqrt{\bigg(\prod_a g_{y_ay_a}(r_{\star})\bigg)}}{\sqrt{g_{xx}(r_{\star})}}E
\ee

The absence of  singular behavior to  observable $J^x$ means the terms under the square-root is regular.  Upon choosing the positive sign, the conductivity is
\be\label{conductivity_one_electric_field}
\sigma=N\f{\sqrt{\bigg(\prod_a g_{y_ay_a}(r_{\star})\bigg)}}{\sqrt{g_{xx}(r_{\star})}}.
\ee  

The solution to the first equation of eq(\ref{conditions_no_density}) gives the desired solution to, $r_{\star}$, as a function of electric field, $E$, and Hawking temperature, $T_H$, as $g_{tt}$ is a function of $T_H$. If we assume that the metric components along the spatial directions are all same then the above formula of conductivity reduces to
\be
\sigma=N \sqrt{\bigg(\prod^{d-2}_1 g_{y_ay_a}(r_{\star})\bigg)}.
\ee
This form of the conductivity is also found in the Maxwell system in \cite{il},  except  the choice of $r_{\star}$, which is $r_{\star}=r_0$.  

Let us find the complete form of the conductivity associated to the Lifshitz metric as written in eq(\ref{lifshitz_geometry}). In which case the relevant equation that gives,  $r_{\star}$,  as a function of electric field, $E$, is
\be\label{alg_equation_rstar}
E^2=r^{2(1+z)}_{\star} \Bigg[1-\bigg(\f{r_0}{r_{\star}}\bigg)^{d+z}\Bigg].
\ee

This algebraic equation is very non-linear in nature and hence   very difficult to find the exact solution,  analytically. However, there exists  exact solutions for few specific cases.  In which case  the number of spatial directions are tied to the exponent $z$ as,  $d=(n-1)z+n$ with $n=0,~1,~2,~3$ and $4$, 
\bea
r_{\star}&=& E^{\f{1}{2(1+z)}},~~~n=0,\nn
r_{\star}&=&\Bigg[\f{(\f{4\pi}{1+z})^{\f{1+z}{z}}T^{\f{1+z}{z}}_H\pm\sqrt{4 E^2+(\f{4\pi}{1+z})^{\f{2(1+z)}{z}}T^{\f{2(1+z)}{z}}_H}}{2}{~~~}\Bigg]^{\f{1}{2(1+z)}},~~~n=1,\nn
r_{\star}&=&\Bigg[E^2+\Bigg(\f{2\pi}{1+z}\Bigg)^{\f{2(1+z)}{z}}T^{\f{2(1+z)}{z}}_H\Bigg]^{\f{1}{2(1+z)}}=\Bigg[E^2+\Bigg(\f{2\pi}{1+z}\Bigg)^{\f{2(1+z)}{z}}T^{\f{2(1+z)}{z}}_H\Bigg]^{\f{1}{d+z}},~~~n=2,\nn
r_{\star}&=&\Bigg[\f{2.3^{\f{1}{3}}E^2+2^{\f{1}{3}}\bigg[9(\f{4\pi}{3(1+z)})^{\f{3(1+z)}{z}}T^{\f{3(1+z)}{z}}_H+\sqrt{81(\f{4\pi}{3(1+z)})^{\f{6(1+z)}{z}}T^{\f{6(1+z)}{z}}_H-12 E^6}\bigg]^{\f{2}{3}}}{6^{2/3}\bigg[9(\f{4\pi}{3(1+z)})^{\f{3(1+z)}{z}}T^{\f{3(1+z)}{z}}_H+\sqrt{81(\f{4\pi}{3(1+z)})^{\f{6(1+z)}{z}}T^{\f{6(1+z)}{z}}_H-12 E^6}\bigg]^{\f{1}{3}}}\Bigg]^{\f{1}{2(1+z)}},~~~n=3,\nn
r_{\star}&=&\Bigg[\f{E^2\pm\sqrt{E^4+4 (\f{\pi}{1+z})^{\f{4(1+z)}{z}}T^{\f{4(1+z)}{z}}_H}}{2}{~}\Bigg]^{\f{1}{2(1+z)}},~~~n=4.
\eea

It is interesting to note that the choice  $n=0$  gives negative exponent $z=-d$, whereas $n=1$ gives $d=1$, which essentially says about a $1+1$ dimensional field theory for any exponent, the choice $n=2,~3$ and $4$ gives the exponent $z=d-2,~\f{d-3}{2}$ and $z=\f{d-4}{3}$, respectively.
   
Now using the spatial part of the metric components from eq(\ref{lifshitz_geometry}) in the expression of current gives, $J\equiv E^{\f{d-1+z}{1+z}}~Y_1$, with the function
\bea\label{y1_specific_case}
Y_1&=&N\Bigg[1+\Bigg(\f{2\pi}{1+z}\Bigg)^{\f{2(1+z)}{z}}\Bigg(\f{T^{1+\f{1}{z}}_H}{E}\Bigg)^2\Bigg]^{\f{d-2}{2(1+z)}}~~~{\rm for}~~~n=2,\nn
Y_1&=&N\Bigg[\f{1\pm\sqrt{1+4 (\f{\pi}{1+z})^{\f{4(1+z)}{z}}\bigg(\f{T^{1+\f{1}{z}}_H}{E}\bigg)^4}}{2}\Bigg]^{\f{d-2}{2(1+z)}}~~~{\rm for}~~~n=4,
\eea
for a couple of cases and  the conductivity in these special cases are
\bea
\sigma&=&N T^{\f{d-2}{z}}_H\Bigg[\Bigg(\f{2\pi}{1+z}\Bigg)^{\f{2(1+z)}{z}}+\Bigg(\f{E}{T^{1+\f{1}{z}}_H}\Bigg)^2\Bigg]^{\f{d-2}{2(1+z)}}~~~{\rm for}~~~n=2,\nn
\sigma&=&\f{N}{2^{\f{d-2}{2(1+z)}}}\Bigg[E^2\pm\sqrt{E^4+4 (\f{\pi}{1+z})^{\f{4(1+z)}{z}}T^{\f{4(1+z)}{z}}_H}{~}\Bigg]^{\f{1}{2(1+z)}}~~~{\rm for}~~~n=4.
\eea

Hence, for very small electric field and at high temperature limit, $E < <T^{1+\f{1}{z}}$, the conductivity follows the  power law behavior, in particular,  $T^{\f{d-2}{z}}_H$.

Let us go away from this special case of $d=(n-1)z+n$ and find the solution 
to $r_{\star}$ from eq(\ref{alg_equation_rstar}). In the weak field limit, $E < <T^{1+\f{1}{z}}$, the solution to $r_{\star}$ can be approximated as
\be
r_{\star}\simeq r_0\Bigg[1+\Bigg(\f{E}{r^{1+z}_0}\Bigg)^2\Bigg]^{\f{d-2}{d+z}}+\cdots=\bigg(\f{4\pi T_H}{d+z}\bigg)^{\f{1}{z}}\Bigg[1+\Bigg(\f{d+z}{4\pi}\Bigg)^{\f{2(1+z)}{z}}\Bigg(\f{E}{T^{1+\f{1}{z}}_H}\Bigg)^2\Bigg]^{\f{1}{d+z}}+\cdots
\ee
which gives the current to leading order 
\be
J_x\simeq N E \bigg(\f{4\pi T_H}{d+z}\bigg)^{\f{d-2}{z}}\Bigg[1+\Bigg(\f{d+z}{4\pi}\Bigg)^{\f{2(1+z)}{z}}\Bigg(\f{E}{T^{1+\f{1}{z}}_H}\Bigg)^2\Bigg]^{\f{d-2}{d+z}}+\cdots,
\ee
whereas in the strong field limit, $E >> T^{1+\f{1}{z}}$, the solution becomes
\be
r_{\star}\simeq E^{\f{1}{1+z}}\Bigg[1+\Bigg(\f{r_0}{E^{\f{1}{1+z}}}\Bigg)^{d+z}\Bigg]^{\f{1}{d+z}}+\cdots=E^{\f{1}{1+z}}\Bigg[1+\bigg(\f{4\pi }{d+z}\bigg)^{\f{d+z}{z}}\Bigg(\f{T^{1+\f{1}{z}}_H}{E}\Bigg)^{\f{d+z}{1+z}}\Bigg]^{\f{1}{d+z}}+\cdots
\ee
which gives the current to leading order 
\be
J_x\simeq N E^{\f{d+z-1}{1+z}} \Bigg[1+\Bigg(\f{4\pi}{d+z}\Bigg)^{\f{d+z}{z}}\Bigg(\f{T^{1+\f{1}{z}}_H}{E}\Bigg)^{\f{d+z}{1+z}}\Bigg]^{\f{d-2}{2(1+z)}}+\cdots.
\ee
This form of current essentially gives us the function  
\be\label{generic_y1}
Y_1=N \Bigg[1+\Bigg(\f{4\pi}{d+z}\Bigg)^{\f{d+z}{z}}\Bigg(\f{T^{1+\f{1}{z}}_H}{E}\Bigg)^{\f{d+z}{1+z}}\Bigg]^{\f{d-2}{2(1+z)}}+\cdots.
\ee

On comparing  this expression of $Y_1$ for $n=2$ case as  in  eq(\ref{y1_specific_case}), it follows that the sub-leading terms to $Y_1$ in eq(\ref{generic_y1}) vanishes exactly for $d=z+2$.

{\bf Multiple electric fields:}\\
Let us consider a situation where we have turned on more than one constant electric field, for simplicity,  let us take the gauge potential  as $A=-(E_1~t+H(r))dx-E_2~t~dy$, which gives the field strength as
\be
F_2=-E_1 dt\w dx-E_2 dt\w dy-H'(r)dr\w dx.
\ee

Let us consider the previous brane configuration, again, but with this new from of the gauge field strength  in the background metric 
\be\label{generic_metric_II}
ds^2_{d+2}=-g_{tt}(r)dr^2+g_{rr}(r)dr^2+g_{xx}(r)dx^2+g_{yy}(r)dy^2+
\sum^{d-2}_{1}g_{ab}(r)dz^adz^b.
\ee

Going through the procedure as outlined above we ended up with 

\be
J^x=\pm N~E_1~\sqrt{\prod^{d-2}_1 g_{z_az_a}(r_{\star})}\sqrt{\f{g_{yy}(r_{\star})}{g_{xx}(r_{\star})}},
\ee
where $r_{\star}$ is to be determined by solving
\be\label{surface_condition_two_electric_fields}
g_{tt}(r_{\star})g_{xx}(r_{\star})=E^2_1+E^2_2 \f{g_{xx}(r_{\star})}{g_{yy}(r_{\star})}.
\ee

Now, note that the functional expression of the current density remains same as is found in the  DBI action with one electric field,  but the condition on $r_{\star}$  
is different. For $g_{xx}(r_{\star})=g_{yy}(r_{\star})$, the condition almost remains the same as for one electric field 
except with the substitution  $E^2_1\rightarrow E^2_1+E^2_2$, but for unequal  $g_{xx}(r_{\star})$ and $g_{yy}(r_{\star})$, one has to find the choice of cutoff $r_{\star}$ by solving  eq(\ref{surface_condition_two_electric_fields}).

{\bf With a constant electric and magnetic field:}\\
Let us re-run the calculation  with a constant electric and  magnetic field, in which case the field strength is
\be\label{frame_b_e_field_strngth}
F_2=-E dt\w dx+B dx\w dy-H'(r)dr\w dx.
\ee

For our choice of constant electric and magnetic field,  the formula of current density becomes 
\be\label{current_b_e_field}
J^x=\pm N~E~\sqrt{\f{\bigg(\prod^{d-2}_{a=1} g_{z_az_a}(r_{\star})g_{yy}(r_{\star})\bigg)}{g_{xx}(r_{\star})}}\sqrt{1-\f{B^2}{E^2}\f{g_{tt}(r_{\star})}{g_{yy}(r_{\star})}}.
\ee

It is easy to note that the current  is  modified with an additional multiplicative factor 
$\sqrt{1-\f{B^2}{E^2}\f{g_{tt}(r_{\star})}{g_{yy}(r_{\star})}}$ in comparison to the cases without any magnetic field. This is because the scale, $r_{\star}$, is also modified by the same multiplicative factor on the right hand side of eq(\ref{conditions_no_density}), but without the square-root.

At first sight, it looks as if the results  of current in eq(\ref{currents_dbi_4d}), after substituting  the solution, are not  compatible with eq(\ref{current_b_e_field}), in $3+1$ dimensions. Actually to compare both the equations we should to go a frame where both the calculation are done in one coordinate system. To do that we can  either do some change of coordinates or directly compute the current using the approach of \cite{kob}. 

In either way note that the computation to eq(\ref{currents_dbi_4d}) is done for which $F^{(\tau)}_{xr}$ vanishes in the frame of $(\tau,~r,~x_i)$. We have used a superscript $(\tau)$ in the expression of field strength to denote  it, which in the frame $(t,~r,~x_i)$ e.g., as in eq(\ref{frame_b_e_field_strngth}) says that they are related as $F^{(\tau)}_{xr}=H'(r)-E/h(r)$. Vanishing of $F^{(\tau)}_{xr}$ means $H'(r)=E/h(r)$, and equating this with the solution to $H'(r)$ that follows from the DBI action,  gives
\be
J^x=\pm N~E,
\ee
 for zero magnetic field. This  precisely  matches  with eq(\ref{currents_dbi_4d}) at the scale, $r=r_{\star}$, up to an  over all  normalization.

\subsection{Subsummary}

To   summarize, in this section, we have  studied  the expression of the current density in terms of one or more constant electric  and magnetic field. Essentially, use of  the prescription of \cite{kob} or equivalently that of eq(\ref{condition_vanishing_energy_static}) and eq(\ref{energy_legendre_transformed_generic}), results in a recipe to calculate the current density in $d+1$ dimensional field theory if the dual bulk geometry  is of the form eq(\ref{generic_metric_II}).  With a constant electric field say  along, $x$, one of the spatial direction, the ratio of the current density to electric field   is the square-root 
of the ratio of the product of metric component (up to an over all factor) of $d-1$ space, which  is perpendicular to $t,~x,~r$ plane, to the metric component along x-axis i.e., eq(\ref{conductivity_one_electric_field}). This quantity should be evaluated at an energy scale, $r_{\star}$, for which  the product of the metric components along $t$ and $x$ axis i.e., $g_{tt}(r_{\star})g_{xx}(r_{\star})$ becomes same as the  square of the electric field i.e., the first equation of eq(\ref{conditions_no_density}). The condition that  determines the point, $r_{\star}$, is generalized when there exists  more than one constant electric field and a constant magnetic field in the theory.

It is worth to emphasize that $r_{\star}$ should be close to the horizon, $r_0$, rather than to boundary because in order for eq(\ref{current_b_e_field}) to make sense. The factor 
$\sqrt{1-\f{B^2}{E^2}\f{g_{tt}(r_{\star})}{g_{yy}(r_{\star})}}$ should be a real quantity and  can happen   only when $r_{\star}$ is close to the horizon for any strength of the magnetic and electric field. This can be seen as follows, close to the horizon the ratio $\f{g_{tt}(r_{\star})}{g_{yy}(r_{\star})}$ is very small whereas close to the boundary this ratio approaches unity.  So for $B > E$ the second factor in the square root can become greater than unity.

\section{With charge density}\label{charge_density}

Let us   discuss the effect of the   non vanishing charge density   along  with the Chern-Simon term on the conductivity. The inclusion of the Chern-Simon term makes an interesting change of the Hall conductivity that is it adds a piece and could potentially change the structure unless we take the axion to be constant.  Moreover, the Chern-Simon term does not make any surprising changes in the Hall angle, $cot~\theta_H=\sigma^{xy}/\sigma^{xx}$ at the leading order in the large density and small magnetic field limit.

\subsection{Charge density with the Chern-Simon term }

It is not {\it a priori}  clear whether  the low energy effective action of the probe brane  admits a Chern-Simon type term or not. We assume that it does and  takes the form similar to that in  string theory except that the target space here is $3+1$ dimensional.
In this section, we have considered the following  form of the field strength
\be\label{generic_field_strength1}
F_2=-E_1 dt\w dx-E_2 dt\w dy+B dx\w dy-H'(r)dr\w dx+h'(r)dr\w dy+\phi'(r)dr\w dt.
\ee
The inclusion of the Chern-Simon term in the probe brane action adds the following  term to the 3+1 dimensional action 
\be
S_{CS}=\mu\int \bigg(\f{[C_0]}{2} F\w F+[C_2]\w F+[C_4]\bigg),
\ee
in the absence of the $B_2$ field from the NS-NS sector. The bulk fields $[C_n]$ are to be understood as the pullback onto the world volume of the probe brane. Let us also assume 
for simplicity, $C_4$ vanishes, the $C_2$ has the following structure, $[C_2]=-{\tilde C}_2(r) dt\w dy$ and $[C_0]$ depends only on the radial coordinate. Using 
the field strength as written in eq(\ref{generic_field_strength1}) results in 
\be
S_{CS}=\mu\int \bigg[C_0(E_1 h'+E_2 H'-B\phi')+{\tilde C}_2 H' \bigg]dt\w dx\w dy\w dr. 
\ee
Let us redefine ${\tilde \mu}:=\mu V_3$, where $V_3$ is the volume of $R^{1,2}$. Finally the Chern-Simon action becomes
\be
S_{CS}={\tilde\mu}\int \bigg[C_0 E_1 h'+(C_0 E_2+{\tilde C}_2) H'-C_0B\phi') \bigg] dr. 
\ee

So, the full action of the probe brane is
\bea
S&=&-{\cal N}\int dr \bigg[g_{rr}(B^2g_{tt}-E^2_1g_{yy}-E^2_2 g_{xx}+g_{tt}g_{xx}g_{yy})+(g_{tt}g_{xx}-E^2_1)h'^2-\nn&&2E_1E_2 h'H'+(g_{tt}g_{yy}-E^2_2)H'^2+2B(E_1h'+E_2H')\phi'-
(g_{xx}g_{yy}+B^2)\phi'^2\bigg]^{1/2}\nn 
&+&
{\tilde\mu}\int dr \bigg[C_0 E_1 h'+(C_0 E_2+{\tilde C}_2) H'-C_0B\phi') \bigg]\nn
&\equiv& -{\cal N}\int dr ~~~{\bar{\cal L}}+{\tilde\mu}\int dr \bigg[C_0 E_1 h'+(C_0 E_2+{\tilde C}_2) H'-C_0B\phi') \bigg]
\eea

Once again the action does not depend on the fields $\phi,~H$ and $h$, so the corresponding momentum are constants. Let us denote the constant momentum for the field $\phi,~h$ and $H$ as $c_{\phi},~c_h$ and $c_H$, respectively
\bea\label{momentum_density}
&&\f{\delta S}{\delta \phi'}\equiv c_{\phi}=-\f{{\cal N}\sqrt{\prod g_{z_az_a}}}{{\bar{\cal L}}}[B(E_1h'+E_2H')-
(g_{xx}g_{yy}+B^2)\phi']-{\tilde \mu}BC_0,\nn
&&\f{\delta S}{\delta h'}\equiv c_{h}=-\f{{\cal N}\sqrt{\prod g_{z_az_a}}}{{\bar{\cal L}}}[(g_{tt}g_{xx}-E^2_1)h'-E_1E_2 H'+BE_1\phi']+{\tilde \mu}E_1C_0,\nn
&&\f{\delta S}{\delta H'}\equiv c_{H}=-\f{{\cal N}\sqrt{\prod g_{z_az_a}}}{{\bar{\cal L}}}[(g_{tt}g_{yy}-E^2_2)H'+BE_2\phi'-E_1E_2 h']+{\tilde \mu}(E_2C_0+{\tilde C}_2]
\eea

From now on, we shall drop the tildes from the field $C_2$ and the coupling $\mu$ so as to avoid cluttering of it. By taking the ratio of the momenta, we determine $h'$ and $\phi'$ in terms of $H'$ 
\bea\label{h_phi_in_terms_of_H_density}
h'&\equiv& \f{h_1}{h_2}, ~~~{\rm where}\nn
h_1&=& g_{yy} \bigg[E_2 (c_H E_1 - c_h E_2-E_1 \mu C_2) g_{xx} + 
    g_{tt} \bigg(B (B c_h + c_{\phi} E_1) + (c_h-E_1 \mu C_0) g_{xx} g_{yy}\bigg)\bigg] H'\nn
h_2&=&
g_{xx} \bigg[E_1 (-c_H E_1 + c_h E_2+E_1 \mu C_2) g_{yy} + 
    g_{tt}\bigg(B (B c_H + c_{\phi} E_2) +\nn && (c_H-E_2 \mu C_0) g_{xx} g_{yy}-\mu C_2 (B^2 + g_{xx} g_{yy})\bigg)\bigg],\nn 
\phi'&\equiv&\f{\phi_1}{\phi_2}, ~~~{\rm where}\nn
\phi_1&=& g_{tt} \bigg[E_1 (B c_h + c_{\phi} E_1) g_{yy} + 
   g_{xx} \bigg(E_2 (B c_H + c_{\phi} E_2)-B E_2 \mu C_2 - (c_{\phi}+B \mu C_0) g_{tt} g_{yy}\bigg)\bigg]H'\nn
\phi_2&=&g_{xx} \bigg[E_1 (-cH E_1 + c_h E_2+E_1 \mu C_2) g_{yy} + 
    g_{tt}\bigg(B (B c_H + c_{\phi} E_2) + \nn&&(c_H-E_2 \mu C_0) g_{xx} g_{yy}-\mu C_2 (B^2 + g_{xx} g_{yy})\bigg)\bigg]
\eea

The function  $H'$ can be evaluated by substituting eq(\ref{h_phi_in_terms_of_H_density}) into the last equation of eq(\ref{momentum_density}). We are not writing down the explicit form of $H'$ as it is a very long expression. As is done previously, the Legendre transformed action 
\bea
S_L&=&S-\int \f{\delta S}{\delta \phi'}\phi'-\int \f{\delta S}{\delta h'}h'-\int \f{\delta S}{\delta H'}H'\nn &=&-{\cal N}\int \f{dr}{\bar{\cal L}} \bigg[g_{rr}(B^2g_{tt}-E^2_1g_{yy}-E^2_2 g_{xx}+g_{tt}g_{xx}g_{yy}) \bigg]\nn
&=&-\int dr \sqrt{\f{g_{rr}}{g_{tt} g_{xx} g_{yy}}}\times\sqrt{[{\cal A_2}(r){\cal A_3}(r)-({\cal A_4}(r))^2]},
\eea
where
\bea\label{A_2-A_3-A_4}
{\cal A_2}(r)&=& B^2 g_{tt} - E^2_2 g_{xx} - E^2_1 g_{yy} + 
g_{tt} g_{xx} g_{yy},\nn
{\cal A_3}(r)&=&  g_{tt} g_{xx} g_{yy} - \bigg(c_H-\mu (E_2 C_0 + C_2)\bigg)^2 g_{xx} -(c_h-E_1 \mu C_0)^2 g_{yy} + (c_{\phi}+B \mu C_0)^2 g_{tt},\nn
{\cal A_4}(r)&=&(c_{\phi}+B \mu C_0) B g_{tt} + (c_h-E_1 \mu C_0) E_1 g_{yy} + \bigg(c_H-\mu (E_2 C_0 + C_2)\bigg) E_2 g_{xx}
\eea  

From now onwards, we shall set $E_2=0$ as it does not change much of the physics that we are going to do. The scale $r_{\star}$ is determined by solving 
\be\label{condition_r_star_density}
{\cal A_2}(r_{\star})=[B^2 g_{tt} -   E^2_1 g_{yy} + 
g_{tt} g_{xx} g_{yy}]_{r_{\star}}=0.
\ee
The form of the currents that follows are
\bea
J^x&=&\mu C_2\mp\f{E_1 g_{yy}}{B^2 + g_{xx} g_{yy}} \sqrt{ 
    (c_{\phi} + B  \mu C_0)^2+ {\cal N}^2 ( B^2 + g_{xx} g_{yy})},\nn
J^y&=&E_1\bigg[\f{  B(c_{\phi}+ B  \mu C_0)}{B^2 + g_{xx} g_{yy}}-\mu C_0\bigg]=\f{E_1 [B c_{\phi}- \mu C_0 g_{xx} g_{yy}]}{B^2 + g_{xx} g_{yy}},
\eea
where we have used eq(\ref{condition_r_star_density}). From which the conductivity follows upon using the Ohm's law 
\be\label{conductivity_with_cs}
\sigma^{xx}=\mp\f{ g_{yy}}{B^2 + g_{xx} g_{yy}} \sqrt{ 
    (c_{\phi} + B  \mu C_0)^2+   {\cal N}^2 (B^2 + g_{xx} g_{yy})},~~~
\sigma^{xy}=\f{B c_{\phi}- \mu C_0 g_{xx} g_{yy}}{B^2 + g_{xx} g_{yy}}.
\ee
The quantity, ${\cal N}^2\sim N^2_f/g^2_s$, where $N_f$ is the number of flavor branes and $g_s$ is the string coupling.
Once again, let us look at a special corner of the parameter space of charge density $c_{\phi}$ and the magnetic field $B$ for which ${\cal N} B$ is very small in comparison to 
density. In this case the conductivity reduces to 
\be
\sigma^{xx}\simeq \mp \bigg[\f{c_{\phi}}{g_{xx}}+{\cal N}^2\f{g_{yy}}{2c^2_{\phi}}+\cdots\bigg],~~~\sigma^{xy}\simeq \f{B c_{\phi}}{g_{xx}g_{yy}}-\mu C_0+\cdots,
\ee   
where the ellipses denote higher powers of magnetic field. Choosing the positive branch from $\sigma^{xx}$ and dropping the second term in $\sigma^{xx}$ gives us
\be
\sigma^{xx}\sim  \f{c_{\phi}}{g_{xx}},~~~\sigma^{xy}\simeq \f{B c_{\phi}}{g_{xx}g_{yy}}-\mu C_0,
\ee  
which in the small $\mu C_0$ limit i.e., $\mu C_0 << \f{B c_{\phi}}{g_{xx}g_{yy}}$, the Hall angle reduces to 
\be
\f{\sigma^{xx}}{\sigma^{xy}}\sim \f{g_{yy}}{B}.
\ee

So the presence of the Chern-Simon term in the action parametrically does not change much of the conductivity in the small magnetic field and large density limit but adds a piece to the off diagonal part of the conductivity. The Hall angle in the high density limit, $\mu C_0 << \f{B c_{\phi}}{g_{xx}g_{yy}}$, remains  same as in the absence of  the Chern-Simon term.

In the presence of a non trivial dilaton, $\Phi$, the form of the conductivities becomes
\bea
\sigma^{xx}&=&\mp\f{ e^{2\Phi}g_{yy}}{B^2 + e^{4\Phi}g_{xx} g_{yy}} \sqrt{ 
    (c_{\phi} + B  \mu C_0)^2+   {\cal N}^2 e^{-2\Phi}(B^2 +e^{4\Phi} g_{xx} g_{yy})}, \nn
\sigma^{xy}&=&\f{B c_{\phi}- \mu C_0 e^{4\Phi}g_{xx} g_{yy}}{B^2 + e^{4\Phi}g_{xx} g_{yy}}.
\eea

Note that we are calculating the conductivities in the Einstein frame, i.e., we have changed the metric components as $g_{ab}\rightarrow e^{2\Phi} g_{ab}$. In the small magnetic field limit, $B\ll e^{2\Phi(r_{\star})}\sqrt{g_{xx}(r_{\star})g_{yy}(r_{\star})}$, and at large density limit, $c_{\phi} \gg B\mu C_0(r_{\star})$, the conductivities    reduces to
\be
\sigma^{xx}\sim \f{e^{-2\Phi}}{g_{xx}}\sqrt{ 
    c^2_{\phi} +   {\cal N}^2 e^{2\Phi} g_{xx} g_{yy}},~~~\sigma^{xy}\sim \f{B c_{\phi}e^{-4\Phi}}{g_{xx}g_{yy}}-\mu C_0.
\ee

If we, further,  take the  axion as constant, $C_0=\theta$, with a rotationally invariant geometry, in the very high density limit,  $c_{\phi} \gg {\cal N} e^{\Phi(r_{\star})} \sqrt{g_{xx}(r_{\star})g_{yy}(r_{\star})}$,  and assume that the first term in the Hall conductivity dominates over the axionic term then there is no solution to dilaton and metric component that can give the result as written in  eq(\ref{expt_result}). However, if we consider a  limit for which,  $Bc_{\phi}e^{-4\Phi(r_{\star})}\ll \mu C_0(r_{\star}) g_{xx}(r_{\star})g_{yy}(r_{\star})$, then the second term in the Hall conductivity dominates over the first term. In which case, it is possible to reproduce  eq(\ref{expt_result}), up to a sign.

Let us consider another  limit,  $c_{\phi} \gg B\mu C_0$ but $c_{\phi} \ll {\cal N} e^{\Phi(r_{\star})} \sqrt{g_{xx}(r_{\star})g_{yy}(r_{\star})}$,  with small magnetic field, in a rotationally invariant geometry with a non constant axion then the conductivities reduces to
\be
\sigma^{xx}\sim {\cal N}  e^{-\Phi},~~~\sigma^{xy}\sim \f{B  c_{\phi}e^{-4\Phi}}{ g^2_{xx} }-\mu C_0.
\ee

Upon comparing with eq(\ref{expt_result}), the dilaton should  go as $e^{\Phi}\sim {\cal N} T$ and  the combination of metric component and axion should be as
\be\label{sol_non_zero_cs}
\f{B  c_{\phi}e^{-4\Phi}}{ g^2_{xx} }-\mu C_0 \sim T^{-3}.
\ee 

It would be interesting to find such background solutions that has the property as is being written in   eq(\ref{sol_non_zero_cs}).

\section{Geometry with two exponents: An example}

In this section we shall write down a gravitational black hole solution for which the geometry exhibits the required two exponents, explicitly. The extremal solution is already found in \cite{ssp} in a specific setting that is with several form field strengths and metric. But to find the non-extremal solution  in that setup is very cumbersome. Instead, here we shall find such solutions by adopting a different form of the gravitational action than that is considered in \cite{ssp}, but it comes up with a cost that is the entropy vanishes even though there is a finite size of the horizon. The on shell action vanishes identically as a result of the vanishing of the free energy and the energy.
Similar kind of behavior was  seen previously in the context of generating Lifshitz type of solutions in \cite{cls} and  \cite{aggh}.

The action that we shall consider is a Ricci squared corrected term to Einstein-Hilbert action  with a cosmological constant
\be\label{action_example}
S= \f{1}{2\kappa^2_4}\int\sqrt{-g}[R-2\Lambda+\beta R^2]\equiv \int L.
\ee
The equation of motion that follows from it is
\be\label{eom_metric}
R_{MN}-\f{1}{2}g_{MN}R+\Lambda g_{MN}+2\beta g_{MN}\Box R-2\beta\nabla_M\nabla_N R+2\beta RR_{MN}-\f{1}{2}\beta R^2 g_{MN}=0.
\ee

The solution to the equation of motion comes as
\be\label{solution_2_exponent}
ds^2=L^2[-r^{2z}f(r)dt^2+r^{2w}dx^2+r^2dy^2+\f{dr^2}{r^2f(r)}],
\ee
which respect the scaling symmetry 
\be
t\rightarrow \lambda^z ~t,~~~x\rightarrow \lambda^w ~x,~~~y\rightarrow \lambda ~y,~~~r\rightarrow \f{r}{\lambda}.
\ee

The function 
\be
f(r)=1-\bigg(\f{r_h}{r} \bigg)^{\alpha_{\pm}},
\ee
where 
\be\label{alpha}
\alpha_{\pm}=1+w+\f{3}{2}z\pm\f{1}{2}\sqrt{-4(1+w^2)+4z+4wz+z^2}.
\ee
From which there follows a restriction on the exponents,
$4z+4wz+z^2 ~\geq~ 4(1+w^2)$ and 
the dimension full objects $\beta$ and $\Lambda$ are
\be
\Lambda=-\f{1}{2L^2}[1+w+z+w^2+z^2+wz],~~~\beta=\f{L^2}{4[1+w+z+w^2+z^2+wz]}
\ee 
with the Hawking temperature 
\be
T_H=\f{\alpha_{\pm}}{4\pi}r^z_h.
\ee

It follows trivially that for a solution with exponents for which $z=1$ and $w=1/2$ satisfies the restrictions that $\alpha_{\pm}$ is a real quantity and hence the solution is real. For this choice of  the exponents the cosmological constant and the coupling are
\be\label{solution_example}
\Lambda=-\f{17}{8L^2},~~~\beta=\f{L^2}{17},~~~\alpha_{\pm}=\f{3\sqrt{2}\pm 1}{\sqrt{2}}.
\ee

If we calculate the entropy of the system using Wald's formula \cite{wald}
\be
S_{BH}=-2\pi \int_{r_h} \f{\p L}{\p R_{abcd}}\epsilon_{ab}\epsilon_{cd},
\ee

where the quantity $\epsilon_{ab}$ is  binormal to the bifurcation surface, and is normalized in such a way that it obeys $\epsilon_{ab}\epsilon^{ab}=-2$. We use the convention of \cite{dg} to calculate it, which reads as
\be
\epsilon_{ab}=\xi_a\eta_b-\xi_b\eta_a,
\ee 
where $\xi$ and $\eta$ are null vectors normal to the bifurcate killing horizon, with $\xi.\eta=1$. In our choice of $3+1$ dimensional metric, the non vanishing components of the null vectors are
\be
\xi_t=-g_{tt}=-L^2r^{2z}f(r),~~~\eta_t=1,~~~\eta_r=-\sqrt{\f{g_{rr}}{g_{tt}}}=-\f{1}{f(r)r^{1+z}}.
\ee
In fact, for the action like eq(\ref{action_example}) the entropy is
\be
S_{BH}=\f{2\pi}{\kappa^2_4}\Bigg(\sqrt{-g} [1+2\beta R]\Bigg)_{r_h},
\ee
and using all the ingredients into this formula gives us zero  entropy, which means the solution eq(\ref{solution_2_exponent}) has the constant curvature: $R=-1/2\beta$. From the trace of the equation of motion to metric eq(\ref{eom_metric}), it follows that the scalar curvature obeys 
\be
R=4\Lambda+6\beta \Box R.
\ee

Now combining these two facts, we obtain the curvature
\be
R=-1/2\beta=4\Lambda,
\ee
which is precisely  the behavior of the solution in eq(\ref{solution_example}).

\subsection{Parameter Space}

In this subsection we shall write down the exact form of both the conductivity and Hall angle that follows from section \ref{charge_density}. Before doing the evaluation of the conductivity we need to know the scale, $r_{\star}$. From eq(\ref{condition_r_star_density}), it follows that for small electric field and magnetic field the scale
\be
 r_{\star}\sim r_h\sim T^{1/z}_H. 
\ee
The correction to this scale occurs in the dimensionless ratios of $E/T^{1+1/z}_H$ and $B/T^{(1+w)/z}_H$.

Now, substituting the explicit form of the metric components from eq(\ref{solution_2_exponent}) into eq(\ref{conductivity_with_cs}) results in
\be
\sigma^{xx}=\mp\f{ r^2_{\star}}{B^2 + r^{2(1+w)}_{\star}} \sqrt{ 
    (c_{\phi} + B  \mu C_0)^2+   {\cal N}^2 \bigg(B^2 + r^{2(1+w)}_{\star}\bigg)},~~~
\sigma^{xy}=\f{B c_{\phi}- \mu C_0 r^{2(1+w)}_{\star}}{B^2 + r^{2(1+w)}_{\star}}.
\ee

In the small magnetic field, large density and at low temperature limit the expression of the  conductivities reduces to
\bea\label{temp_transport}
&&\sigma^{xx}\sim c_{\phi} r^{-2w}_{\star}+\cdots\sim c_{\phi}T^{-2w/z}_{H},~~~
\sigma^{xy}\sim B c_{\phi}r^{-2(1+w)}_{\star}- \mu C_0+\cdots\sim B c_{\phi}T^{-2(1+w)/z}_{H}- \mu C_0,\nn &&\Longrightarrow \sigma^{xx}/\sigma^{xy}\sim r^{2}_{\star}/c_{\phi}+\cdots\sim (T^{2/z}_{H})/c_{\phi}+\cdots,
\eea
where in the last line we have assumed $B c_{\phi} > \mu C_0 T^{2(1+w)/z}_H$.  
Demanding that this temperature dependence of conductivities should match the experimental results, eq(\ref{expt_result}),
gives us the following values of exponent, $z=1$, and $w=1/2$. So, the above form of the exponents gives us the strange metal behavior of copper-oxide systems as  seen in experiments \cite{cwo},~\cite{tm},~\cite{rc}. If we consider the other regime of parameter space with a constant axion at low temperature for which the magnetic field is small in comparison to charge density such that, $B c_{\phi} < \mu C_0 T^{2(1+w)/z}_H$, then the off diagonal conductivity does not depend on the temperature for constant axion. So, is not of much interest as far as the experimental results are concerned. Hence, this regime of parameter space may not be that useful. However, if we consider  the non constant axion field in the same limit, i.e.,  $B c_{\phi} < \mu C_0 T^{2(1+w)/z}_H$, then by matching with  eq(\ref{expt_result}), we get the exponents as $2w=z$ and the axion field should have the following behavior, $C_0 \sim T^{-3}_H \sim r^{-3 z}_{\star} $. It would be interesting to find such background solutions. 

\subsection{Fermi Liquid}

In this subsection we shall reproduce a well known transport  properties of the Fermi liquid theory. It is known, see for example  \cite{gs}, that  the conductivity at low temperature goes as $\sigma_{FL}\sim T^{-2}$. Now upon using eq(\ref{temp_transport}), we see that in order to reproduce this particular behavior of the temperature requires us to take the exponents as $w=z$. Here the exponents are not fixed to a particular value. 
In the next section, we shall  demand that the specific heat  should have a linear dependence of temperature, parametrically. The result of  this,  fixes the exponents to $z=w=-2$. Note,  for this choice of exponents the quantity $\alpha_{\pm}$ defined    in eq(\ref{alpha}) becomes pure imaginary, which is an artifact of the action we used to construct it.  However, in what follows we shall not be worried  
about the nature of $\alpha_{\pm}$ as we believe the above mentioned constraint on the exponents can be removed by looking at better solutions.

\section{Probe brane thermodynamics}

In this section we shall study some  thermodynamic properties of the probe brane but without the Chern-Simon term and non trivial dilaton. Let us recall that the charge carriers are introduced via probe brane and the study of their thermodynamic behavior  is very important   so as to have a better understanding of the nature of quantum critical point. It is reported in \cite{at} and \cite{german_group}, for a review see \cite{gs} that at low temperature the specific heat (for  NFL ) goes as $C_V\sim T~Log~T$. But unfortunately, with our choice of exponents as demanded by the transport properties:  $z=1$, and $w=1/2$, gives us the specific heat to go instead as $C_V\sim T^3_H$. This kind of behavior of specific heat resembles that of the Debye theory.

Let us see this particular behavior of specific heat in detail. We shall proceed to calculate the free energy of the probe brane system following \cite{kss}. The proper holographic  treatment is also done in \cite{kkp} and  \cite{kob1}. The Gibbs free energy, i.e., the thermodynamic potential, $\Omega$, in the grand canonical ensemble is just the negative of the on shell value of the action times temperature. Here, we have chosen to work in the  canonical ensemble. The easiest way to include  the effect of the charge density and magnetic field is by using  the field strength, $F_2=\phi'(r)dr\w dt+B dx\w dy$ in the DBI action.  In $3+1$ dimensions, using the metric as in eq(\ref{generic_metric_II}) gives us the thermodynamic potential and chemical potential $ \mu=\int^{\infty}_{r_h} dr F_{rt}=\int^{\infty}_{r_h}dr \phi'$. The chemical  potential, $\mu$, should not be confused with the Chern-Simon coupling   that appeared in section 3.2.

\be
\Omega={\cal N} V_2 \int^{\infty}_{r_h} dr \f{(g_{xx}g_{yy}+B^2)\sqrt{g_{tt}g_{rr}}}{\sqrt{g_{xx}g_{yy}+B^2+\rho^2}},~~~
\mu=\rho\int^{\infty}_{r_h} dr \f{\sqrt{g_{tt}g_{rr}}}{\sqrt{g_{xx}g_{yy}+B^2+\rho^2}},
\ee
where ${\cal N}\rho=c_{\phi}$, and $c_{\phi}$ is the charge density. $V_2$ is the flat space volume of $x,~y$ plane. Using the metric structure as written in eq(\ref{solution_2_exponent}) gives 
\be
\Omega={\cal N} V_2 \int^{\infty}_{r_h} dr \f{(r^{2+2w}+B^2)r^{z-1}}{\sqrt{r^{2+2w}+B^2+\rho^2}},~~~
\mu=\rho\int^{\infty}_{r_h} dr \f{r^{z-1}}{\sqrt{r^{2+2w}+B^2+\rho^2}}.
\ee

For generic choice of the exponents, the integral in the  thermodynamic potential and in the chemical potential diverges at UV, so we need to regulate it. The way we shall do is to subtract an equivalent amount but without the charge density and magnetic field. It means 
\bea
\Omega&=&{\cal N} V_2 \int^{\infty}_{0} dr \bigg(\f{r^{1+2w+z}}{\sqrt{r^{2+2w}+B^2+\rho^2}}-r^{z+w}\bigg)-{\cal N} V_2 \int^{r_h}_{0} dr \f{r^{1+2w+z}}{\sqrt{r^{2+2w}+B^2+\rho^2}} \nn
&+& {\cal N} V_2B^2\int^{\infty}_{0} dr \bigg(\f{r^{z-1}}{\sqrt{r^{2+2w}+B^2+\rho^2}}-r^{z-w-2}\bigg)-{\cal N} V_2B^2\int^{r_h}_{0} dr \f{r^{z-1}}{\sqrt{r^{2+2w}+B^2+\rho^2}},\nn
\mu&=&\rho\int^{\infty}_{0} dr \bigg(\f{r^{z-1}}{\sqrt{r^{2+2w}+B^2+\rho^2}}-r^{z-w-2}\bigg)-\rho\int^{r_h}_{0} dr \f{r^{z-1}}{\sqrt{r^{2+2w}+B^2+\rho^2}}
\eea

The second term in the square bracket of the first equation should not be there when $z+w=0$. Similarly, the second term in the square bracket of the first equation comes into picture only when  $z > 2+w$, so also for the second term in the second square bracket. Let us assume  the case, where $z \not> 2+w$ and $z\neq -w$. It means, we want to regulate it in the following way
\bea
\Omega&=&{\cal N} V_2 \int^{\infty}_{0} dr \bigg(\f{r^{1+2w+z}}{\sqrt{r^{2+2w}+B^2+\rho^2}}-r^{z+w}\bigg)-{\cal N} V_2 \int^{r_h}_{0} dr \f{r^{1+2w+z}}{\sqrt{r^{2+2w}+B^2+\rho^2}} \nn
&+& {\cal N} V_2B^2\int^{\infty}_{0} dr \f{r^{z-1}}{\sqrt{r^{2+2w}+B^2+\rho^2}}-{\cal N} V_2B^2\int^{r_h}_{0} dr \f{r^{z-1}}{\sqrt{r^{2+2w}+B^2+\rho^2}},\nn
\mu&=&\rho\int^{\infty}_{0} dr \f{r^{z-1}}{\sqrt{r^{2+2w}+B^2+\rho^2}}-\rho\int^{r_h}_{0} dr \f{r^{z-1}}{\sqrt{r^{2+2w}+B^2+\rho^2}}
\eea

After doing these integrals, we find
\bea\label{grand_potential_mu}
&&\f{\Omega}{{\cal N} V_2}= \alpha(w,z)~(B^2+\rho^2)^{\f{z+w+1}{2+2w}}- \f{1}{(2+2w+z)}~\f{r^{(2+2w+z)}_h}{\sqrt{B^2+\rho^2}}\times\nn &&{}_2F_1[1+\f{z}{2+2w},\f{1}{2};2+\f{z}{2+2w};-\f{r^{2+2w}_h}{B^2+\rho^2}]+\f{B^2}{z\sqrt{\pi}}~(B^2+\rho^2)^{\f{z-w-1}{2+2w}}~\Gamma\bigg(\f{1+w-z}{2+2w}\bigg)\nn && \Gamma\bigg(\f{2+2w+z}{2+2w}\bigg)-\f{r^z_h}{z}\f{B^2\rho}{\sqrt{B^2+\rho^2}}
~{}_2F_1\Bigg[\f{z}{2+2w},~\f{1}{2};~1+\f{z}{2+2w};~-\f{r^{2+2w}_h}{B^2+\rho^2}\Bigg],\nn
&&\mu=\f{1}{z\sqrt{\pi}}~(B^2+\rho^2)^{\f{z-w-1}{2+2w}}~\Gamma\bigg(\f{1+w-z}{2+2w}\bigg)~\Gamma\bigg(\f{2+2w+z}{2+2w}\bigg)-\nn&&\f{r^z_h}{z}\f{\rho}{\sqrt{B^2+\rho^2}}
~{}_2F_1\Bigg[\f{z}{2+2w},~\f{1}{2};~1+\f{z}{2+2w};~-\f{r^{2+2w}_h}{B^2+\rho^2}\Bigg],\nn
\eea
where $\alpha(w,z)$ is a function of the exponents, whose explicit structure is not that important for the understanding of  thermodynamics. $\Gamma(x)$ and ${}_2F_1[a,b;c;x]$ are the gamma function and  hypergeometric function, respectively. In the limit of high density, low magnetic field and low temperature, i.e., $T^{\f{1+w}{z}}/\sqrt{B^2+\rho^2}~\ll ~1$,  eq(\ref{grand_potential_mu}) can be expanded in the series form. The  free energy in the canonical ensemble, $F=\Omega+\mu J^t$, where $J^t={\cal N} V_2\rho$ is the charge. From this the entropy density goes as
\be
s=-\f{1}{V_2}\bigg(\f{\p F}{\p T_H} \bigg)=s_0+\f{{\cal N}}{2z\sqrt{B^2+\rho^2}}~\bigg(4\pi/\alpha_{\pm} \bigg)^{\f{2+2w+z}{z}}T^{\f{2+2w}{z}}_H,
\ee
where $s_0=\f{4\pi{\cal N}}{z\alpha_{\pm}}\sqrt{B^2+\rho^2}$ is the entropy density at zero temperature. The specific heat is defined as the heat capacity per unit volume, and at low temperature, it goes as
\be
C_V=T_H\bigg(\f{\p s}{\p T_H} \bigg)= \f{{\cal N}}{\sqrt{B^2+\rho^2}}~\bigg(\f{1+w}{z^2}\bigg)~\bigg(4\pi/\alpha_{\pm} \bigg)^{\f{2+2w+z}{z}}T^{\f{2+2w}{z}}_H.
\ee

The magnetic susceptibility, which we shall call as  susceptibility, at low temperature
\be
\chi/V_2=-\bigg(\f{\p^2 F}{\p B^2} \bigg)=-\chi_0(B,\rho)+\f{{\cal N}4\pi}{z\alpha_{\pm}}\f{\rho^2}{(B^2+\rho^2)^{3/2}}T_H,
\ee
where $\chi_0$ is some function of $B$ and $\rho$, whose exact form is not that illuminating.  The effect of the Chern-Simon term with the field strength, $F_2=\phi'(r)dr\w dt+B dx\w dy$, is to replace $\rho$ in all of the above formulas   by $\rho+\mu\theta B$, where we have considered the axion field to be a constant and identified it with $C_0\equiv \theta$.  

\subsection{At high temperature, low magnetic field and low density}

In this subsection, we shall write down the behavior of thermodynamic quantities in the high temperature but low magnetic field limit. One of the main reason to study  this regime of parameter space is to see  the behavior of  susceptibility. Probably, it is correct to say that when we are in the proximity of quantum critical point the magnetization   should not obey the Curie-Weiss type behavior in the high temperature limit.

The temperature dependence of free energy in this regime can be obtained very easily by looking at the following integrals 
\bea
\f{\Omega}{{\cal N} V_2}&\sim&- \int^{r_h}_{0} dr \f{r^{1+2w+z}}{\sqrt{r^{2+2w}+B^2+\rho^2}} -B^2\int^{r_h}_{0} dr \f{r^{z-1}}{\sqrt{r^{2+2w}+B^2+\rho^2}},\nn
&=&-\f{ r^{z+w+1}_h}{z+w+1}-\f{B^2 r^{z-w-1}_h}{z-w-1}+\f{ (B^2+\rho^2)}{2(z-w-1)}r^{z-w-1}_h+\f{B^2 (B^2+\rho^2)}{2(z-3w-3)}r^{z-3w-3}_h+\cdots,\nn
\mu&\sim&-\rho\int^{r_h}_{0} dr \f{r^{z-1}}{\sqrt{r^{2+2w}+B^2+\rho^2}}=-\f{\rho r^{z-w-1}_h}{z-w-1}+\f{\rho (B^2+\rho^2)}{2(z-3w-3)}r^{z-3w-3}_h+\cdots.
\eea

So, the Free energy in the canonical ensemble has the following behavior in the high temperature limit
\bea
\f{F}{{\cal N} V_2}&=&-\f{1}{z+w+1}\bigg(4\pi/\alpha_{\pm} \bigg)^{\f{z+w+1}{z}}T^{\f{z+w+1}{z}}_H-\f{(B^2+\rho^2)}{2(z-w-1)}\bigg(4\pi/\alpha_{\pm} \bigg)^{\f{z-w-1}{z}}T^{\f{z-w-1}{z}}_H+\nn
&& \f{(B^2+\rho^2)^2}{2(z-3w-3)}\bigg(4\pi/\alpha_{\pm} \bigg)^{\f{z-3w-3}{z}}T^{\f{z-3w-3}{z}}_H+\cdots.
\eea

From which the magnetization, $\f{M}{{\cal N} V_2}=-\bigg(\f{\p F/({\cal N} V_2)}{\p B} \bigg)$ and the susceptibility,  $\f{\chi}{{\cal N} V_2}=-\bigg(\f{\p^2 F/({\cal N} V_2)}{\p B^2} \bigg)$ are
\bea
\f{M}{{\cal N} V_2}&=&-\f{B}{(z-w-1)}\bigg(4\pi/\alpha_{\pm} \bigg)^{\f{z-w-1}{z}}T^{\f{z-w-1}{z}}_H-\f{2B(B^2+\rho^2)}{(z-3w-3)}\bigg(4\pi/\alpha_{\pm} \bigg)^{\f{z-3w-3}{z}}T^{\f{z-3w-3}{z}}_H,\nn
\f{\chi}{{\cal N} V_2}&=&\f{1}{(z-w-1)}\bigg(4\pi/\alpha_{\pm} \bigg)^{\f{z-w-1}{z}}T^{\f{z-w-1}{z}}_H-\f{2(3B^2+\rho^2)}{(z-3w-3)}\bigg(4\pi/\alpha_{\pm} \bigg)^{\f{z-3w-3}{z}}T^{\f{z-3w-3}{z}}_H\nn
&\equiv&{\widetilde \chi}_0 T^{\f{z-w-1}{z}}_H-{\widetilde \chi}_1T^{\f{z-3w-3}{z}}_H
\eea

Now, if we demand that the magnetization or more precisely, the susceptibility has the Curie-Weiss type behavior, the above result forces us to put the following constraints on the exponents
\be\label{curie_type}
2z=1+w.
\ee 

Recalling the results to the exponents that follows from the study of conductivity and Hall angle in section 4, suggests that near to quantum critical point the system does not show the Curie-Weiss type behavior. In fact the behavior of susceptibility using the exponents, $z=1$, and $w=1/2$, gives 
\be
\chi={\widetilde \chi}_0 ~T^{-1/2}_H,
\ee
and for Fermi liquid, $z=-2=w$, at high temperature limit goes as
\be
\chi={\widetilde \chi}_0 ~T^{1/2}_H.
\ee

The Curie-Weiss type behavior is possible only when eq(\ref{curie_type}) is obeyed. From which it follows trivially that the asymptotically AdS spacetime  possesses such kind of behavior,  as an example,  for which  $z=1=w$. 
Once again the effect of the Chern-Simon term is to replace  $\rho$ in all of the above formulas   by $\rho+\mu\theta B$, for constant axion as stated in the previous section.

\section{Conclusion}

In this paper we have shown that there exists  two possible ways, (see eq(\ref{summary_eq_1})), with different symmetries to find the precise  temperature dependence of the  longitudinal conductivity and Hall angle, $1/T$ and $T^2$, respectively as seen in  the non-Fermi liquid.  The  calculation is done similar in spirit to the proposal of \cite{kob}, where the charge density is introduced via flavor brane. It is done  in a generic $3+1$ dimensional (bulk)  background solution possessing the  symmetries, like scaling,  time translation, spatial translations.  The result of the calculation suggests   that in order to get the desired experimental result for the transport quantities that is mentioned above, in the high density limit, we should not take the spatial part of the metric components  same, i.e., $g_{xx}\neq g_{yy}$. It means the theory should have the symmetries like scaling, time translation, and spatial translation symmetry but without any rotational symmetry. For this purpose, we have considered a metric with two exponents, $z$ and $w$, as defined in eq(\ref{two_exponent_scaling}).  The end result of this requirement is that the exponents take the values, $z=1,$ and $w=1/2$.

The study of the thermodynamic behavior of various physical quantities are   equally important in the study of the  quantum critical point or otherwise. For the above choice of the exponents, the specific heat  at low temperature goes as $C_V\sim T^3_H$, which resembles that of  the Debye type. The susceptibility at zero magnetic field and at low temperature goes as $\chi=-\chi_0+{\rm (constants)}\times T_H/\rho$, where $\chi_0$ is a function of charge density.

However, if we consider a theory to have the symmetries like pseudo-scaling (non trivial scalar field), time translation, spatial translation and rotation  in the low density limit, we can reproduce eq(\ref{expt_result}) without the need to introduce two exponents. 
We leave the detailed study of the thermodynamic behavior of this class of solution for future research.

From this study, there follows an interesting outcome: we are completely ruling out those background solutions that possesses the symmetries like scaling symmetry, time translation, spatial translation and rotational symmetry. In other words these symmetries are not consistent with eq(\ref{expt_result}). 

The  transport and thermodynamic behavior of various physical  quantities at high density and at low temperature  can be summarized in this two exponent model  as follows :
\be
    \begin{tabular}{ | l | l | l | l |l |p{3cm} |}
    \hline
    Type & Physical quantity & Expt. result & Ref & In this model & Experimental result forces the choice of Exponents \\ \hline
NFL & Conductivity & $T^{-1}$ & \cite{cwo},\cite{tm}, \cite{rc} & $T^{-2w/z}_H$ & $z=1,~w=1/2$ \\ \hline
NFL & Hall Angle & $T^{2}$ & \cite{cwo},\cite{tm}, \cite{rc} & $T^{2/z}_H$ & $z=1,~w=1/2$ \\ \hline\hline
FL & Conductivity & $T^{-2}$ & \cite{gs} & $T^{-2w/z}_H$ & $z=-2,~w=-2$ \\ \hline
FL & Hall Angle & Not known &  & $T^{2/z}_H\sim T^{-1}_H$ &  \\ 
& & to author & & &\\\hline
\end{tabular}
\ee 
and 
\be
    \begin{tabular}{ | l | l | l | l |l |p{3cm} |}
    \hline
    Type & Physical quantity & Expt. result & Ref & In this model & Experimental result forces  \\ 
& & & & &the choice of Exponents \\\hline
FL & Specific heat & $T$ & \cite{gs} & $-T^{(2+2w)/z}_H$ & $z=-2,~w=-2$ \\ \hline
FL & Susceptibility:  & independent  & \cite{gs} & $-\chi_0+{\rm const/\alpha_{\pm}}$ & $z=-2,~w=-2$ \\ 
 & $\chi(B=0) $ & of $T$ & & $\times T_H/\rho$  & \\ \hline\hline
NFL & Specific heat & should not  & \cite{gs} & $T^{(2+2w)/z}_H\sim T^3_H$ &  \\ 

& & be as $T$ & & \\\hline
NFL & Susceptibility:  & Not known  &  & $-\chi_0+$ &  \\ 
 & $\chi(B=0) $ &  to author & & ${\rm const}\times T_H/\rho$  & \\ \hline\hline
\end{tabular}
\ee 

At high temperature, low density and low magnetic field  limit

\be\label{high_temp_limit}
    \begin{tabular}{ | l | l | l  |l |l |}
    \hline
    Type & Physical quantity & Exponents &  In this model & Prediction \\ \hline
NFL & Specific heat & $z=1,~w=1/2$ &  $T^{(1+w)/z}_H$ & $T^{3/2}_H$ \\ \hline
NFL & Susceptibility  & $z=1,~w=1/2$   & $T^{(z-w-1)/z}_H$ & $T^{-1/2}_H$ \\ \hline\hline
FL & Specific heat & $z=-2=w$ &  $ T^{(1+w)/z}_H$ & $T^{1/2}_H$ \\ \hline
FL & Susceptibility  & $z=-2=w$ &  $ T^{(z-w-1)/z}_H$ & $T^{1/2}_H$ \\ \hline\hline
AdS & Specific heat  & $z=1=w$ &  $T^{(1+w)/z}_H$ & $T^2_H$ \\ 
Spacetime & & & &\\ \hline
AdS & Susceptibility  & $z=1=w$ &  $T^{(z-w-1)/z}_H$ & $T^{-1}_H$ \\
Spacetime & & & &  \\\hline\hline
\end{tabular}
\ee 

From eq(\ref{high_temp_limit}), it follows that it's only the asymptotically AdS spacetime that shows  the Curie-Weiss type of behavior. From which, it is natural to think that the asymptotically AdS spacetime may be associated to metals, more specifically to the  paramagnets, even though the specific heat shows a quadratic dependence of temperature.

In this study, we have constructed  a background black hole geometry with two exponents, for illustration. The future goal would be to construct other background solutions having non trivial  spacetime thermodynamics, i.e., the thermodynamics of adjoint degrees of freedom, in the sense of having non zero entropy for finite horizon size and may be  non zero  free energy, depending on the requirement of the model, which is {\it a priori} not clear at present. Moreover, the thermodynamic quantities in the Fermi liquid phase need to be real.

There are several other checks that needed to be done. In particular,  the AC conductivity, $\sigma(\omega)$, which in the interval $T_H < ~\omega < ~{\tilde \Omega} $, shows a very specific behavior \cite{dvdm}, where $\omega$  and ${\tilde \Omega} $ are the frequency and some high energy cutoff scale, respectively. This result of \cite{dvdm} for copper oxide systems puts some serious restrictions on the form of the bulk geometry. In the study of superconductors \cite{hr} at low temperature (close to extremality), it was suggested that if the potential energy close to IR behaves as, $V=V_0/r^2$, then the real part of AC conductivity goes as, $\Re[\sigma(\omega)]\propto \omega^{\sqrt{4V_0+1}-1}$. Now upon matching with the results of \cite{dvdm},  we get  $V_0=-2/9$, i.e., there should be an attractive  potential energy close to IR. This is an interesting prediction but we leave this aspect of holographic model building for future research.

There is one further comment that deserve to be mentioned. In \cite{hpst}, it is shown that both for scaling and presudo-scaling theories with unbroken rotational symmetry in the $x,~y$ plane, the resistivity and the AC conductivity have the following temperature and frequency dependence, at low temperature,   $\rho\sim T^{\nu_1}$, and $\sigma(\omega)\sim \omega^{-\nu_1}$ for $\nu_1 \leq 1$. Now if we demand eq(\ref{expt_result}) on this result, then it fixes $\nu_1=1$, it means, $\sigma(\omega)\sim \omega^{-1}$, which is not allowed by \cite{dvdm}. So, it is natural to think of some more exotic models that are either proposed in \cite{hpst} or that  discussed in this paper, in order to get as close as possible  to the experimental results.

\section{Acknowledgment}
It is a pleasure to thank Ofer Aharony 
for raising the issue of lattice structure and to Rene Meyer for some useful discussions. Thanks are to  the members of  CQUeST for  their help.

This work was supported by the Korea Science and Engineering Foundation (KOSEF) grant funded by the Korea
government (MEST) through the Center for Quantum Spacetime (CQUeST) of Sogang University with Grant No, R11-2005-021.

\section{Appendix A: Solution to Maxwell system}

In this section we shall write down the exact solution to the  Maxwell's equation of motion in the notation of \cite{bkls}. Let us start with a system whose dynamics is  described by Maxwell's action 
\be
S=-\f{1}{4}\int d^{d+1}x \f{\sqrt{-g}}{g^2_{YM}}F_{MN}F^{MN},
\ee
with coordinate dependent coupling, $g_{YM}$, whose explicit dependence we do  not specify. Also, we assume that the $d+1$ dimensional spacetime possesses  the symmetries like rotation in $d-1$ dimensional space, along with time and spatial translations has a structure like
\be\label{EF-metric}
ds^2_{d+1}=-h(r)d\tau^2+2 d\tau dr+e^{2s(r)}\delta_{ij}dx^idx^j,
\ee
where the radial coordinate can have a range, $ r_0 \leq r \leq r_c $, with $r_0$ denote the horizon of a black hole and $r_c$ is the upper cutoff, which is the UV. The non-vanishing components of the field strength's are $F_{\tau r},~F_{\tau i},~F_{ij},~F_{ir}$.

In terms which the equations of motion are
\bea\label{eom-maxwell}
&&e^{2s}\p_{\tau}F_{\tau r}+\p^i F_{\tau i}=h \p^i F_{ir},\nn
&& \p_{\tau}F_{ir}+[(2-p)s'-\phi']F_{\tau i}-\p_r F_{\tau i}+h' F_{ir}=-h\p_r F_{ir}+h[(2-p)s'-\phi']F_{ir}+e^{-2s}\p^jF_{ji},\nn
&&\p_r F_{\tau r}+(ps'+\phi')F_{\tau r}+e^{-2s}\p^i F_{ir}=0,
\eea
where we have used $1/g^2_{YM}=e^{\phi(r)}$ and in the derivatives the indices $i,~j,\cdots$ are raised using $\delta^{ij}$ i.e., $\p^i=\delta^{ij}\p_j$.  The normalization of the coupling is assumed to be $\phi(r_0)=0$. We shall solve these equations of motion along with the Bianchi identities 
\bea\label{bianchi_identity}
&&\p_{\tau}F_{ri}+\p_rF_{i\tau}+\p_i F_{\tau r}=0,\nn
&&\p_{\tau}F_{ij}+\p_iF_{j\tau}+\p_j F_{\tau i}=0,\nn
&&\p_{r}F_{ij}+\p_iF_{jr}+\p_j F_{ ri}=0,
\eea
with the in falling boundary condition at the horizon, which means the momentum flux tangent to the horizon vanishes i.e.,
$T_{rr}(r_0)=0$ \cite{bkls}, which means $F_{ir}(r_0)=0$. 

The current and charge density at the horizon are
\be
J_i(\tau,~x_i,r_0)=F_{i\tau}(\tau,~x_i,r_0)~~~~q(\tau,~x_i,r_0)=F_{r\tau}(\tau,~x_i,r_0),
\ee
which obey the continuity equation $\p_{\tau}q+\p^iJ_i=0$, courtesy the first equation of eq(\ref{eom-maxwell}) after setting the condition $s(r_0)=0$.

\subsection{Exact Solutions}

In this subsection, we shall find the exact solution to the Maxwell system, first, for $3+1$ dimensional space time and then for any arbitrary spacetime dimension. 

3+1 dimension:\\

Let us denote the spacetime coordinate as $\tau,~x,~y$ and $r$. The solution for which the coupling is constant i.e., $\phi'=0$ with a non-trivial electric field and a constant magnetic field 
\be\label{exact_solution_3+1_zero_density}
F_{xr}=0,~~~F_{yr}=0,~~~F_{r\tau}=0 ,~~~F_{y\tau}=E_y(\tau),~~~
F_{xy}={\rm constant}\equiv B,~~~F_{x\tau}=E_x(\tau),
\ee
for some functions $E_x(\tau)$ and $E_y(\tau)$ whose functional form is not fixed by the equations of motion or the Bianchi identity. \\

For $p=2$, there exists another exact solution for which the 
the coupling is constant i.e., $\phi'=0$ and the rest of the components of  field strength are
\be
F_{xr}=0,~~~F_{yr}=0,~~~F_{r\tau}=q~ e^{2s(r)},~~~F_{y\tau}=E_y(\tau),~~~
F_{xy}={\rm constant}\equiv B,~~~F_{x\tau}=E_x(\tau),
\ee
where $q$ is a constant and the functions $E_x(\tau)$ and $E_y(\tau)$ whose functional form is not fixed by the equations of motion or the Bianchi identity.\\

Any arbitrary dimension:\\

There exists  exact solution to the Maxwell system in any arbitrary spacetime dimension but unfortunately with zero electric field. In fact all the components of field strength vanishes
except 
\be
F_{r\tau}=q e^{-ps(r)-\phi(r)},
\ee
where $q$ is a constant. One can find another solution with a non-trivial electric field provided the inverse coupling goes as
\be
1/g^2_{YM}=e^{-(p-2)s(r)+{\rm constant}},
\ee
with  $F_{r \tau }=q e^{-2s(r)}$ and the other  non-vanishing component of the field strength is
\be
F_{i\tau}=E_i(\tau)
\ee
for some functions $E_i(\tau)$, again whose functional form is not fixed by the equations of motion or the Bianchi identity.

\section{Appendix B: Energy Minimization}
 After extremizing eq(\ref{energy_electric_field}), it follows that the extremum occurs when the following equation is satisfied at some $r$

\bea\label{energy_minimization_electric_field}
&&\bigg[g'_{rr}(g_{tt}g_{xx}-E^2)+g_{rr}(g'_{tt}g_{xx}+g_{tt}g'_{xx})\bigg]\bigg[N^2(\prod g_{y_ay_a})-\f{c^2}{g_{tt}}\bigg]+\nn&&\bigg[g_{rr}(g_{tt}g_{xx}-E^2)\bigg]\bigg[N^2(\prod g_{y_ay_a})'+\f{c^2g'_{tt}}{g^2_{tt}}\bigg]=0.
\eea 
 Let us denote it as $r_m$ which is a function of $(T_H,~E,~J^x)$. But recall that vanishing and reality of energy, $H_L$, implies that it occurs at a scale $r_{\star}$, which is a function of $T_H$ and $E$ only. So one can  ask the question: Can $r_{\star}$ be same as $r_m$ i.e., 
$r_{\star}(T_H,~E)=r_m(T_H,~E,~J^x)$? The answer to this question is: It can happen only if the current $J^x$ is a function of $T_H$ and $E$. If we take the case as in eq(\ref{conditions_no_density}) then it gives us a solution to eq(\ref{energy_minimization_electric_field}). In fact for this solution the energy extremization at $r_{\star}$ again is in the indeterminate form i.e., $\bigg(\f{dH_L}{dr}\bigg)_{r_{\star}}=\f{0}{0}$ because in this case both $H_L$ and the numerator of $\f{dH_L}{dr}$ vanishes. So we shall take the physical reason of choosing a scale $r_{\star}$ is the condition of reality and vanishing of energy, $H_L$. In general, a priori, it is not clear what other value of $r$ one should choose so as to find the current as a function of temperature and electric field that solves eq(\ref{energy_minimization_electric_field}) for which $r_{\star}=r_m$.

\end{document}